\documentclass[twocolumn,superscriptaddress,showpacs,amsmath,amssymb,prl,longbibliography,nofootinbib]{revtex4-1}

\usepackage{listings}
\usepackage{dcolumn}  
\usepackage{bm}       
\usepackage[usenames,dvipsnames]{color}
\definecolor{URLCOL}{rgb}{0,0.52,0.83} 
\definecolor{LINKCOL}{rgb}{0.05,0.5,0} 
\definecolor{orange}{rgb}{0.6,0.3,0} 
\definecolor{CITECOL}{rgb}{0.25,0,0.48} 
\usepackage{epstopdf}

\usepackage[pdftex,bookmarks,breaklinks,bookmarksopen,bookmarksnumbered,colorlinks,linkcolor=LINKCOL,linktocpage,citecolor=CITECOL,urlcolor=URLCOL,pdfpagemode=UseOutline,pdftex]{hyperref}

\usepackage{booktabs}
\usepackage{comment}

\definecolor{TITLECOL}{rgb}{0.1,0.2,0.7} 
\definecolor{SECOL}{rgb}{0.1,0.2,0.7} 
\definecolor{CONTENTSCOL}{rgb}{0.1,0.2,0.7} 
\definecolor{SSECOL}{rgb}{0.25,0,0.48} 
\definecolor{SSSECOL}{rgb}{0.2,0.08,0.53} 
\definecolor{FINCOL}{rgb}{0.01,0.3,0.07} 

\def\check#1{{\color{orange}#1}} 

\definecolor{URLCOL}{rgb}{0,0.17,0.43} 
\definecolor{LINKCOL}{rgb}{0.05,0.4,0} 
\definecolor{CITECOL}{rgb}{0.35,0,0.48} 

\def\bea{\begin{eqnarray}}
\def\eea{\end{eqnarray}}
\def\ben{\begin{equation}}

\def\een{\end{equation}}
\def\benu{\begin{enumerate}}
\def\enu{\end{enumerate}}
\def\bei{\begin{itemize}}
\def\eei{\end{itemize}}

\def\br{{\bf r}}

\def\n{n}

\def\sss{\scriptscriptstyle\rm}

\def\g{_\gamma}

\def\x{_{\sss X}}
\def\c{_{\sss C}}
\def\s{_{\sss S}}
\def\xc{_{\sss XC}}

\def\Hxc{_{\sss HXC}}

\def\H{_{\sss H}}

\def\ee{_{\rm ee}}

\def\wt{\texttt{w}}
\def\b{^\tau}

\def\t{^{\tau}}

\def\tl{^{\tau,\lambda}}
\def\g{_\gamma}

\begin{document}

\title{
Thermal density functional theory:  Time-dependent linear response and 
approximate functionals from the fluctuation-dissipation theorem
}
\author{Aurora Pribram-Jones}
\affiliation{Lawrence Livermore National Laboratory, Livermore, CA 94550}
\affiliation{Department of Chemistry, University of California, Berkeley, CA 94720}
\author{Paul E. Grabowski}
\affiliation{Department of Physics and Astronomy, University of California, Irvine, CA 92697}
\author{Kieron Burke}
\affiliation{Department of Chemistry, University of California, Irvine, CA 92697}
\affiliation{Department of Physics and Astronomy, University of California, Irvine, CA 92697}
\date{\today}

\begin{abstract}
The van Leeuwen proof of linear-response 
time-dependent density functional theory (TDDFT) is generalized to
thermal ensembles.  This allows generalization
to finite temperatures of the Gross-Kohn relation,
the exchange-correlation kernel of TDDFT, 
and fluctuation dissipation theorem for DFT.  
This produces a natural method for generating new thermal
exchange-correlation (XC) approximations.
\end{abstract}


\maketitle

Kohn-Sham density functional theory (KS-DFT) is a popular and well-established approach
to electronic structure problems in many areas, especially materials science and chemistry\cite{FNM03}.
The Kohn-Sham method imagines a fictitious system of non-interacting fermions with the
same density as the real system\cite{KS65} and from which the ground-state energy can be extracted.
Only a small fraction of the total energy, called the exchange-correlation (XC) energy,
need be approximated to solve any ground-state electronic problem\cite{FNM03}, and modern approximations
usually produce sufficient accuracy to be useful\cite{PGB15}.
The advent of TDDFT generalized this method to time-dependent problems\cite{RG84}.  Limiting TDDFT to linear-response
yields a method for extracting electronic excitations\cite{C96,PGG96},
once another functional, the XC
kernel, is also approximated.

But there is growing interest in systems in which the electrons are not close to zero temperature.
Warm dense matter (WDM) is partially ionized, solid-density matter having a temperature near
the Fermi energy. 
It has wide-ranging applications including
the astrophysics of giant planets and white dwarf atmospheres\cite{NHKFR08,LHR09,KDLMF12,MH13,WM12,WM12b,MW10,WM10},
cheap and ultra-compact particle accelerators and
radiation sources\cite{T14,C13,BFBKF12}, and
the eventual production of clean, abundant
energy via inertial confinement fusion\cite{NWTZ72,CFM73}.
One of the most successful methods for simulating equilibrium warm dense matter combines
DFT\cite{HK64,KS65} and molecular dynamics\cite{CP85} to capture quantum mechanical effects
of WDM electrons and the classical behavior of ions\cite{NHKFR08,LHR09,KDLMF12,MH13,WM12,WM12b,MW10,WM10,DKC02,KCHLC10,DM12}.
Such simulations use the Mermin theorem\cite{M65} to generate a KS scheme at finite
temperature, defined to generate the equilibrium density and free energy.
In practice, the XC free energy is almost always approximated with a ground-state
approximation, but formulas for thermal corrections are being developed\cite{SD14,KSDT14,PD84,TI86,DAC86}.

Many processes of interest involve perturbing an equilibrium system with some
time-dependent (TD) perturbation, such as a laser field\cite{RA00} or a rapidly moving nucleus as in 
stopping power\cite{FBCR10,FBBBB13,ZFGLC15}.
Of great interest within the WDM community are calculations of spectra, dynamic structure factors, 
and the flow of energy between electrons and ions\cite{HBNST06,BSCKS12,CVG13,VG14}.  
Spectra expose a material's response to excitation by electromagnetic radiation, which would
facilitate experimental design and analysis.  Dynamic structure factors can be
related to the x-ray scattering response, which is being developed as
a temperature and structural diagnostic tool for WDM\cite{GR09}.
Thus it would appear that a TD version of the Mermin formalism is required.
A theorem is proven in Li et al.\cite{LL85,LT85}, but the formalism assumes the
temperature is fixed throughout the process, and so cannot describe e.g., equilibration
between electrons and ions.  Moreover,
the proof requires the Taylor expansion of the perturbing potential as a function
of time, just as in the Runge-Gross (RG) theorem\cite{RG84}.  This can be problematic
for initial states with cusps\cite{YMB12}, such as at the nuclear centers. (Recent efforts\cite{L01,RL11} have focused on avoiding these complications
at zero temperature.)
Finally, the RG proof requires
invocation of a boundary condition to complete the
one-to-one correspondence between
density and potential\cite{GK90}, which create subtleties when 
applied to extended systems\cite{MSB03}.

In the present work, we prove the RG
theorem at finite temperature within linear response by generalizing the
elegant linear response proof of van Leeuwen\cite{L01} to thermal ensembles.
Our proof avoids several of the
drawbacks mentioned above, while still providing a solid grounding to 
much of WDM theoretical work.  We then define the exchange-correlation kernel at
finite temperature and generalize the Gross-Kohn equation.  Finally, we extend
the fluctuation-dissipation theorem of ground-state DFT to finite temperatures,
and show how this provides a route to {\em equilibrium} free energy XC approximations.

Consider a system of electrons in thermal and particle equilibrium with a bath
at some temperature, $\tau$, and with static equilibrium density $\n\b(\br)$.
The system extends throughout space with a finite average density, i.e., the
thermodynamic limit has been taken.  The limit of isolated atoms or
molecules is achieved by then taking the separation between certain nuclei to infinity.
In this sense, no surface boundary condition need be invoked\cite{GK90},
as the density never quite vanishes, while 
the average particle number per atom or molecule molecule is finite.
These electrons are perturbed at $t=0$
by a potential $\delta v(\br,t)$ that is Laplace-transformable.
To avoid complex questions of equilibration, we consider only the
linear response of the system, so that the perturbation does not affect
the temperature of the system as, e.g., Joule heating is a higher
order effect\cite{KL57}.
The Kubo response formula for the density change in response to $\delta v$
is
\ben
\delta\n\b(\br,s) = \int d^3r'\, \chi\b(\br,\br',s)\, \delta v(\br',s),
\label{dn}
\een
where the Laplace transform
\ben
\delta v(\br,s) = \int_{0}^\infty dt\, e^{-st}\, \delta v(\br,t)
\een
is assumed to exist for all $s>0$.
Within the grand canonical ensemble\cite{PPGB14,CDK15}, the equilibrium 
density-density response function is\cite{SL13}:
\ben
\chi\b(\br,\br',s)=
i\sum_{ij}\, \wt_i \frac{\Delta n^{\tau *}_{ij}(\br)\Delta \n\t_{ij}(\br')}
{s-i\omega_{ji}} + c.c.,
\een
where
\ben
\Delta \n\t_{ij}(\br) = \langle i | \hat \n(\br) | j \rangle - \delta_{ij}\
\n\b(\br)
\label{eqn:Dn}
\een
are matrix elements of the density fluctuation operator. The energy-ordered indices $i,j$ run over all many-body states (both bound and 
continuum\cite{ABBW09}) with all particle numbers,
but $\Delta \n\t_{ij}$ vanishes unless $N_i=N_j$. The transition frequencies 
$\omega_{ji}=E_j-E_i$, and the statistical weights $\wt_i$  are thermal occupations 
for the equilibrium statistical operator 
$\hat{\Gamma}^\tau=\sum_{i}\wt_i |\Psi_i\rangle\langle \Psi_i |$ and obey
$\wt_i<\wt_j$ if $E_i>E_j$ and $N_i=N_j$. This condition is satisfied by the grand canonical
ensemble of common interest with
$\wt_i=\exp[-(E_i-\mu N_i)/\tau]/\sum_{i}\exp[-(E_i-\mu N_i)/\tau]$.

We also need the (Laplace-transformed) one-body potential
operator:
\ben
\delta \hat V(s) = \int d^3r\, \hat \n(\br)\, \delta v(\br,s),
\een
and its matrix elements:
\ben
\delta V_{ij}(s) = \langle i | \delta\hat V(s) | j \rangle.
\label{eqn:matel}
\een
Its expectation value is
\ben
\delta V\b(s) = \sum_i \wt_i\, V_{ii}(s)
= \int d^3r\, \n\b(r)\, \delta v(\br,s),
\een
so that matrix elements of its fluctuations are
\ben
\Delta V\t_{ij}(s) = \delta V_{ij}(s)- \delta_{ij}\, \delta V\b(s).
\een
Then consider the expectation value:
\ben
m\b(s) = \int d^3r\, \delta\n\b(\br,s)\, \delta v(\br,s).
\een
Inserting Eq. (\ref{dn}) and using the definitions, we find
\ben
m\b(s) = 
-\sum_{ij}\wt_i \mid\Delta V\t_{ij}(s)\mid^2
\frac{2\omega_{ji}}{s^2+\omega_{ji}^2}.
\een
This is rearranged as
\ben
\label{eqn:pos}
m\b(s)=-2\sum_{i=0}^\infty\sum_{j=i+1}^\infty
\frac{(\wt_i-\wt_j)\omega_{ji}}{s^2+\omega_{ji}^2}\mid\Delta V\t_{ij}(s)\mid^2.
\een
We have ordered all states by energy regardless of particle number here for simplicity, though this is not strictly necessary since different particle number subsystems do not interact. 
For now, we assume no degeneracies.
Then the above expression, $m\b(s)$, vanishes
only if
every 
$\Delta V\t_{ij}(s)$ does for $i\neq j$ because of our assumption that
$(\wt_i-\wt_j)\omega_{ji}>0$ if $i\ne j$. 

The usual statement of the RG theorem is that no two potentials
that differ by more than an inconsequential function of time alone can give
rise to the same density (for fixed statistics, interparticle interaction, and 
initial state\cite{RG84}).   Imagine two such perturbations exist, yielding
the same density response.  
Since, in linear response, the density response is proportional to the perturbation, we can
subtract one from the other, and the statement to be proved is that
there is no non-trivial perturbation with zero density response.
If it did exist, then $m\b(s)$ would vanish and our algebra shows that
every $\Delta V\t_{ij}(s)$ with $i\ne j$ would also. Finally,
\ben
\sum_{k=1}^{N_i}\delta v(\br_k,s)
\Psi_j(\br_1\ldots\br_{N_i})=\sum_i \delta 
V_{ij}(s) \Psi_i(\br_1\ldots\br_{N_i}),
\label{eqn:tdep}
\een
which can be proven by integrating over all coordinates with $\Psi_k^*$.
Then, as $\Delta V\t_{ij}(s) = \delta V_{ij}(s)$ for $i\neq j$,
and must vanish if there is no density response, the sum on the right of
Eq. (\ref{eqn:tdep}) collapses to just the $j$-th term, showing that
$\delta v(\br,s)$ must be spatially independent.

We can also include a finite number ($M$) of degenerate excited eigenstates.  (For the complications involved
when the ground-state is degenerate, see Ref. \cite{G15}).
For such states, $\omega_{ij}=0$ and the argument above no longer
implies $\delta V_{ij}(s)$ vanishes, as the perturbation couples degenerate states within the same subspace. But simply choose at least $M$
points in the $3N$-dimensional coordinate space that are not on any
nodal hypersurface of the degenerate subspace.  Then the only solution to Eq. (\ref{eqn:tdep})
is again that $\delta v (\br,s)$ must be independent of $\br$.
%

Thus we have generalized the
van Leeuwen proof to thermal ensembles, even with finite degeneracies
among excited states.
Our proof applies to {\em any} ensemble with 
weights that monotonically decrease with increasing energy for
each particle number\cite{PYTB14,G16}. This avoids complications caused by cusps in initial 
wavefunctions\cite{YMB12,YB13}. 
Extension to spatially periodic potentials is straightforward, as no
boundary condition\cite{GK90} was invoked\cite{MSB03}.

In order for the above result to be of practical use, we consider the KS scheme
for finite-temperature, time-dependent systems and provide a
method for generating XC approximations.
We assume the equilbrium Mermin-Kohn-Sham (MKS)\cite{M65,KS65} potential exists.  At this point,
we switch to using the more familiar Fourier-transform notation, but in fact
all results and definitions apply only to Laplace-transformable perturbations. (In practice, this distinction rarely matters, but occasional formal difficulties
arise if this restriction is not made, see Ref. \cite{MK87} and Sec. 3.2 of Ref. \cite{L01}.)
First
we generalize the Gross-Kohn response formula\cite{GK85} to thermal ensembles. Define
\begin{align}
\chi\b(\br,\br',\omega)&=\notag\\
\sum_{jk}\, \wt_j &\left\{\frac{\langle j | \hat \n(\br) | k \rangle \langle k | \hat \n(\br') | j \rangle}
{\omega-\omega_{kj}+i\eta} - \frac{\langle j | \hat \n(\br') | k \rangle \langle k | \hat \n(\br) | j \rangle}
{\omega+\omega_{kj}+i\eta}\right\},
\end{align}
where $\eta\to 0^+$\cite{Yc88}.

Because of our proof of one-to-one correspondence,
we can invert the response function (excluding a constant), and write
\ben
\left(\chi\t\right)^{-1}(12) = \frac{\delta v(1)}{\delta \n(2)},
\label{chiinv}
\een
where $1$ denotes the coordinates $\br,t$, and 2 another pair\cite{KBP94}.
The standard definition of XC is:
\ben
v\s(1) = v(1) + v\H(1) + v\xc(1),
\een
where $v\s$ is the one-body potential of the non-interacting KS system and $v\H$ is the Hartree potential\cite{BW12}. Differentiating with respect to $\n(2)$, this yields
\ben
\left(\chi\t\s\right)^{-1}(12)=\left(\chi\t\right)^{-1}(12) + f\H(12) + f\xc\t(12),
\label{kernel}
\een
which defines the XC kernel at finite temperature, where 
$\chi\s\t$ is the KS response function\cite{Yc88} and the traditionally defined Hartree
contribution is simply
\ben
f\H(12)=\frac{\delta(t_1-t_2)}{\mid \br_1-\br_2 \mid}.
\een
This follows the definition within the Mermin formalism\cite{M65}
(but see Refs. \cite{PPGB14} and \cite{PYTB14} for alternative choices and their consequences).
Inverting yields the thermal Gross-Kohn equation\cite{GK85}:
\ben
\chi\t(12)=\chi\s\t(12) + \int d3d4\, \chi\s\t(13) f\Hxc\t(34) \chi\t(42).
\label{thGK}
\een
A simple approximation is then
the thermal adiabatic local
density approximation (thALDA), in which
the thermal XC kernel is approximated using the XC free energy density per particle
for a finite-temperature uniform gas, $a\xc^{\tau,{\rm unif}}$:
\ben
f\xc^{\tau,{\rm thALDA}}[\n](\br,\br',\omega) = 
\frac{d^2 a\xc^{\tau,{\rm unif}}(\n)}{d^2 \n}\Bigg|_{\n(\br)}\, \delta(\br-\br'),
\label{thALDA}
\een
which ignores its nonlocality in space and time, and could be used to generalize
ALDA calculations of excitations in metals and their surfaces\cite{Liebsch97}.

We next deduce the fluctuation-dissipation theorem for
MKS thermal DFT calculations. This allows us to connect the response function and the Coulomb interaction
through the dynamical structure factor\cite{LP77}.
In the MKS scheme, the XC contributions to the free energy are defined via
\begin{align}
A\t[n]&=T[n]+V\ee[n]+V[n]-\tau S[n]\\
&=T\s[n]+U[n]+V[n]-\tau S\s[n]+A\xc\t[n].
\end{align}
By subtraction,
\ben
A\t\xc[n]=T\t\xc[n]+U\xc\t[n]-\tau S\xc\t[n]
\een
where $T$ denotes kinetic, $U$ denotes potential, and $S$ entropic components.
Using many-body theory, the density-density response function determines the
potential contribution to correlation\cite{FW71,U11}, just as in the ground
state\cite{D12}:
\begin{align}
U\c\t&=&V\ee[\hat{\Gamma}^\tau[n]] - V\ee[\hat{\Gamma}\s^\tau[n]]~~~~~~~~~~~~\\
&=&-\int d\br \int d\br' \int_0^\infty \frac{d\omega}{2\pi}\coth{\bigg(\frac{\omega}{2\tau}\bigg)}  
\frac{\Im \Delta \chi\t(\br,\br',\omega)}{|\br-\br'|},
\label{Uct}
\end{align}
where $\Delta\chi\t=\chi\t-\chi\s\t$.
By introducing a coupling-constant $\lambda$
while keeping the density fixed, the thermal connection formula\cite{PB15}
yields
\begin{align}
&A\c\t[\n]=\notag\\
&\lim_{\tau''\to\infty}\frac{\tau}{2} \int_\tau^{\tau''} \frac{d\tau'}{\tau'^2}\, 
\int d\br\int d\br'\int \frac{d\omega}{2\pi}\coth{\bigg(\frac{\omega}{2\tau}\bigg)} \frac{\Im\Delta\chi^{\tau'}[\n\g](\br,\br',\omega)}
{|\br-\br'|},
\label{AcFD}
\end{align}
where the scaled density is $\n\g(\br) = \gamma^3 \n(\gamma\br)$ and $\gamma=\sqrt{\tau'/\tau}$.
This is exact, but only if the exact thermal XC kernel is used, as defined by
Eq. (\ref{kernel}).  If the kernel
is omitted, the result is the thermal random-phase approximation\cite{BWG05}.

Next, we discuss the many applications of Eq. (\ref{AcFD}).
There has been tremendous progress in implementing and testing the random phase approximation
for calculating the XC energy in ground-state calculations and such calculations, while
more expensive than standard DFT, are becoming routine\cite{EYF10,EF11,PRRS12}.  
Our results provide a thermal
generalization that could likewise be used to generate new thermal XC approximations
for equilibrium WDM calculations.
At finite temperature, the XC hole
fails to satisfy the simple sum rules\cite{PK98b} that have proven
so powerful in constructing ground-state
approximations\cite{BPW97}.   But our formula uses
instead the XC kernel. Inserting Eqs (\ref{thGK},\ref{thALDA}) into Eq. (\ref{AcFD}) yields
thALDA-RPA, a new approximation to the equilibrium correlation energy, that
can be applied to any system.
Another, simpler approximation is ALDA, in which
only the zero-temperature XC energy is used in the kernel.  Both can be
relatively easily evaluated for a uniform gas, and the resulting $a\xc\t(r\s)$
found from  Eq. (\ref{AcFD}) compared with an accurate parametrization\cite{KSDT14}.
Even in the uniform gas, thALDA is an approximation because both the $q$- and
$\omega$-dependence of the true $f\xc\t$
are missing; thus the efficacy of these approximations can be tested on the uniform case.

Next we discuss which known exact conditions on the zero-temperature kernel apply to the thermal kernel,
and which do not.  Because the equilibrium solution is a
minimum of the thermal free-energy functional,
the zero-force theorem\cite{U11} 
\ben
\int d^3 r\int d^3 r' \n\t(\br)\n\t(\br') f\xc\t(\br,\br',\omega)=0
\een
should be satisfied and
the kernel should be symmetric in its
spatial arguments. However, any simple formula for a one-electron system\cite{WYB12} is not
true at finite temperature, as the particle number is only an average in the grand canonical ensemble\cite{PK98b,CDK15}.

A last set of conditions is found by considering the coupling-constant
dependence in DFT.  A parameter $\lambda$ is introduced that multiplies the
electron-electron interaction, while keeping the density constant.  Because of
simple scaling relations, the $\lambda$-dependence can be shown to be
determined entirely by coordinate scaling of the density as in Eq. (\ref{AcFD}), i.e.,
determined by the functional itself, evaluated at different densities.
This is used in both ground-state DFT\cite{LP85} and in time-dependent
DFT\cite{HPB99}, and has been generalized to the thermal
case\cite{PPFS11,PB15}.  Although the thermal connection formula does not
require this relation for the response function, it is useful in many contexts.
From the Lehmann representation\cite{SL13} of $\chi\t$\cite{FW71}, we find
the $\lambda$-dependent response function satisfies:
\ben
\chi\tl[\n](\br,\br',\omega)=\lambda^4\, \chi^{\tau/\lambda^2}[n_{1/\lambda}](\lambda\br,\lambda\br',\omega/\lambda^2).
\label{chitl}
\een
Insertion into the definition of $f\xc$ yields:
\ben
f\xc\tl[\n](\br,\br',\omega)=\lambda^2\, f\xc^{\tau/\lambda^2}[n_{1/\lambda}](\lambda\br,\lambda\br',\omega/\lambda^2),
\label{fxctl}
\een
and the potential perturbation scales as:
\ben
\delta v\xc\tl[\n](\br,\omega)=\lambda^2\, \delta v\xc^{\tau/\lambda^2}[\n_{1/\lambda}](\lambda\br,\omega/\lambda^2).
\een
Insertion of the scaling relation for the
kernel into the thermal connection formula yields a more familiar
analog to the ground-state formula.

The exchange kernel must scale linearly with coupling constant, so Eq. (\ref{fxctl}) produces a rule
for scaling of the exchange kernel:
\ben
f\x\t[\n\g](\br,\br',\omega)=\gamma 
f\x^{\tau/\gamma^2}[n](\gamma\br,\gamma\br',\omega/\gamma^2).
\label{fxtg}
\een
Because the poles in $f\xc$ are $\lambda$-dependent, we expect pathologies similar to those in
zero-temperature TDDFT if 
the exact frequency-dependent $f\x\t$ is used in Eq. (\ref{AcFD})\cite{HB08}.
But adiabatic EXX (AEXX), not including frequency-dependence, produces a well-defined
approximation to the thermal free energy in which the kernel is non-local.  
This and the other proposed approximations above could prove useful
in WDM simulations when thermal XC effects are relevant (but see \cite{SPB15} for discussion of the subtleties involved in thermal XC approximations).

In conclusion, we have generalized the proofs and constructions of
TDDFT within the linear response formalism to thermal
ensembles, including those containing a finite number of degeneracies.
We have avoided ambiguities about the relative
perturbative and thermal equilibration time scales,
allowed for degenerate excited states more common in finite-temperature ensembles, avoided invoking boundary conditions and the requirement of 
Taylor expandability, and provided firm footing for  finite-temperature,
time-dependent KS-DFT in the linear response regime.
Definition of relevant KS quantities led to description of their properties under scaling.  
Further, we have shown that these quantities, in combination with
the thermal connection formula, produce new routes
to thermal DFT approximations for use in equilibrium MKS calculations.
Implementation and tests of these approximations is ongoing.

APJ acknowledges support from DE-FG02-97ER25308 and the University of California President's Postdoctoral Fellowship, PG from DE14-017426, and KB
from CHE-1464795 NSF.  Part of this work was performed under the auspices of the U.S. Department of Energy by Lawrence Livermore National Laboratory under Contract DE-AC52-07NA27344.

\label{page:end}
\end{document}